\documentclass[article,accept,moreauthors,pdftex,10pt,a4paper,preprints]{mdpi} 

\firstpage{1} 

\history{}

\usepackage[utf8]{inputenc}
\usepackage[T1]{fontenc}
\usepackage{hyperref}
\usepackage{graphicx,bm,amsmath,color,scalerel}
\usepackage{bbold}
\usepackage{wasysym}
\usepackage{verbatim}
\usepackage{physics}
\usepackage{comment}
\usepackage{subcaption}
\graphicspath{ {./img/} }
\usepackage{nomencl}
\makenomenclature
\usepackage{etoolbox}
\usepackage{cancel}
\renewcommand\nomgroup[1]{%
  \item[\bfseries
  \ifstrequal{#1}{Z}{Abbreviations}{%
  \ifstrequal{#1}{N}{Number Sets}{%
  \ifstrequal{#1}{O}{Other Symbols}{}}}%
]}

\newcommand{\be}{\begin{equation}}
\newcommand{\ee}{\end{equation}}
\newcommand{\beq}{\begin{eqnarray}}
\newcommand{\eeq}{\end{eqnarray}}
\newcommand{\ba}{\begin{align}}
\newcommand{\ea}{\end{align}}

\Title{Curved Momentum Space, Locality, and Generalized Space-Time}

\Author{José Manuel Carmona$^{1}$*\orcidA{}, José Luis Cortés$^{1}$\orcidB{} and José Javier Relancio$^{1}$\orcidC{}}

\AuthorNames{José Manuel Carmona, José Luis Cortés and José Javier Relancio}

\address{%
$^{1}$ \quad Departamento de F\'{\i}sica Te\'orica and Centro de Astropartículas y Física de Altas Energías (CAPA),
Universidad de Zaragoza, Zaragoza 50009, Spain; jcarmona@unizar.es (J.M.C); cortes@unizar.es (J.L.C.); relancio@unizar.es (J.J.R.)}

\corres{Correspondence: jcarmona@unizar.es}

\abstract{We establish the correspondence between two apparently unrelated but in fact complementary approaches of a relativistic deformed kinematics: the geometric properties of momentum space and the loss of absolute locality in canonical spacetime, which can be restored with the introduction of a generalized spacetime. This correspondence is made explicit for the case of $\kappa$-Poincaré kinematics and compared with its properties in the Hopf algebra framework.}


\begin{document}

\section{Introduction}
\label{sec:introduction}
Before renormalization was discovered, there were other proposals to avoid the ultraviolet divergences in quantum field theory (QFT). In the 1930s, Born~\cite{Born:1938} considered that, since there is a ``reciprocity'' (name chosen from the lattice theory of crystals) between space-time and momentum variables (for example, in a plane wave), there might also be a curved momentum space in analogy to the curved space-time proposed in general relativity (GR). Moreover, he found that, as a consequence of a curvature in momentum space, a ``quantized'' (noncommutative) space-time appears in a natural way. This idea was discussed also by Snyder some years after~\cite{Snyder:1946qz}. In 1947, Snyder showed the first example of a noncommutative space-time, thinking that perhaps a lattice structure for space-time could be a key ingredient missing in QFT. 

However, when renormalization was established, these ideas were forgotten. In the last decades, a quantum gravity theory (QGT) has been searched for due to the inconsistencies between GR and quantum theory (QT). Some attempts to formulate a QGT at a fundamental level include string theory~\cite{Mukhi:2011zz,Aharony:1999ks,Dienes:1996du}, loop quantum gravity~\cite{Sahlmann:2010zf,Dupuis:2012yw}, supergravity~\cite{VanNieuwenhuizen:1981ae,Taylor:1983su}, or causal set theory~\cite{Wallden:2010sh,Wallden:2013kka,Henson:2006kf}. In most of them, a minimum length and a noncommutative space-time appear in a natural way. However, the main problem of these theories is the lack of observable phenomenology. 

A completely different approach has been carried out in the so-called doubly special relativity (DSR) theories that intend to be a low-energy limit of a QGT that could have some experimental observations (see Reference~\cite{AmelinoCamelia:2008qg} for a review). In this context, the usual kinematics of special relativity (SR) is deformed while maintaining a relativity principle, so that a new relativistic deformed kinematics (RDK) appears: there is a deformed dispersion relation (DDR) and a deformed conservation law (DCL) for momenta, and in order to have a relativity principle, there are deformed Lorentz transformations of the momenta (DLT) making the previous ingredients compatible. Usually, these theories are constructed through a mathematical tool called Hopf algebras~\cite{Majid:1995qg}, and the most common example used in the literature is $\kappa$-Poincaré~\cite{Majid1994,Lukierski1995}.

The possible connection between a curved momentum space and an RDK has been suggested in many papers, from the point of view of groups~\cite{KowalskiGlikman:2002ft} and from the fundamental ingredients of the geometry~\cite{AmelinoCamelia:2011bm,Lobo:2016blj}. 
In Reference~\cite{AmelinoCamelia:2011bm}, there are two independent geometric entities that define the kinematics: the dispersion relation is obtained as the square of the distance in momentum space from the origin to a point, while the composition law is related to a non-affine connection. In Reference~\cite{Lobo:2016blj}, the only geometric ingredient is the metric in momentum space, and the authors checked that, 
for the particular case of $\kappa$-Poincaré in the bicrossproduct basis~\cite{KowalskiGlikman:2002jr}, the composition law defines isometries of a de Sitter momentum metric; however, they did not find a way to deduce this fact. Moreover, in both papers, there is a lack of understanding of how to implement a relativity principle, i.e., some deformed Lorentz transformations that make the dispersion relation and the composition law compatible.

In Reference~\cite{Carmona:2019fwf}, it was proposed another way to understand the deformed kinematics from the nontrivial curvature of a momentum space. While the dispersion relation is also given by the square of the distance in momentum space, the composition law and the Lorentz transformations in the one-particle system are given by the ten isometries of a maximally symmetric four-dimensional momentum space. The isometries leaving the origin invariant are just the Lorentz transformations of the one-particle system in the RDK, while the remaining four isometries (translations in momentum space) are related to the deformed conservation laws of momenta in the RDK.   Whereas the Lorentz isometry generators close a particular algebra (Lorentz), this is not the case for the generators of translations, which form a 10-dimensional Lie algebra with the Lorentz generators but not in an unambiguous way. Moreover, the two-particle Lorentz transformations can also be deduced in this scheme~\cite{Carmona:2019fwf}, allowing one to keep a relativity principle in the deformed kinematics.  

An alternative to the geometric perspective of an RDK is based on the (non)locality of interactions. In Reference~\cite{AmelinoCamelia:2011bm}, it was shown that, since translations are defined by the total momentum, which is deformed due to the nonlinear conservation law, interactions are only local for observers placed at the interaction point but not for any other translated observer. This effect was baptized as relative locality, which differs from the absolute locality that characterizes the space-time of SR. 

Moreover, in the Hopf algebra scheme, the coproduct of momenta (which defines the composition law) leads to a noncommutative space-time through the pairing operation~\cite{KowalskiGlikman:2002jr}. This association is carried out by mathematical procedures, without any mention to physical arguments. However, in~\cite{Carmona:2017cry,Carmona:2019vsh}, it was shown that it is possible to implement a locality of interactions in a generalized space-time, for which the coordinates do not commute. In fact, it was found in~\cite{Carmona:2019vsh} that there is a restriction on the possible kinematics that allows one to have local interactions. It was also proven that the generalized space-time that implements locality coincides with the noncommutative space-time obtained in Hopf algebras.   

While the geometry of momentum space and a generalized space-time based on the implementation of locality seem to be different unrelated perspectives, they have an RDK as a common ingredient, and then there should be a relation between them. In this paper, we establish this connection, showing the analogies of both frameworks and how they complement to each other. 

In Section~\ref{sec:DRI}, we introduce the ingredients of a relativistic deformation of the special relativistic (SR) kinematics. In Section~\ref{sec:geometry}, we explain our understanding of an RDK from the geometrical point of view, showing how to obtain all their ingredients through a maximally symmetric momentum space. In Section~\ref{sec:locality}, we see that a nonlocality of interactions appears in the  canonical space-time variables and how locality can be recovered for a particular choice of coordinates, which in fact do not commute. In \mbox{Section~\ref{sec:complementarity}}, we compare both frameworks and establish analogies between them. 
 
\section{Deformed Relativistic Invariance}
\label{sec:DRI}
Let us start by specifying what we mean by a relativistic deformed kinematics (RDK). It is defined by a composition law of momenta, $\oplus$, which is a mapping 
\be
\oplus: \:\: {\cal M} \otimes {\cal M} \to {\cal M},
\ee
where ${\cal M}$ is the momentum space. Given two points with coordinates $p$ and  $q$ in momentum space, the composition law defines a new point $\oplus(p,q)\doteq p\oplus q$. The coordinates $p$ and $q$ are identified with the momenta of two particles. 

The system of two particles with momenta $p$ and $q$ can have two different values for the total momentum ${\cal P}$, ${\cal P}=p\oplus q$ or ${\cal P}=q\oplus p$. Note that the composition law is such that $\oplus(p,q)\neq \oplus(q,p)$; otherwise, one does not have a deformation of the relativistic kinematics but just a different choice of momentum variables in the SR kinematics. 
On the other hand, if the composition law is associative, then the composition of momenta in a multiparticle system is determined by the composition law of two momenta. The discussion of the case of a nonassociative composition law needs additional prescriptions for the definition of the possible momenta of a system of more than two particles, but if they are defined by successive compositions, then the relativistic invariance of multiparticle kinematics is guaranteed by the relativistic invariance of the two-particle system.

The composition law $\oplus$ defines an RDK when one can identify the following: 
\begin{itemize}
\item a representation of the Lorentz transformations ($J$) in momentum space 
\be
J_\omega: \:\: {\cal M} \to {\cal M} \quad\quad\quad k'_\mu = [J_\omega(k)]_\mu \doteq J_\mu(\omega, k),
\ee
where $\omega_{\mu\nu}$ is the six parameters of a general Lorentz transformation and $J_\mu(\omega, k)$ is nonlinear functions of the momentum coordinates $k_\mu$, which define a nonlinear representation of the Lorentz group in momentum space. 
\item another representation of the Lorentz group of transformations ($J^{(2)}$) in the system of two particles such that, when the total momentum is ${\cal P}=p\oplus q$,
\be
J^{(2)}_\omega: \quad {\cal M} \otimes {\cal M} \to {\cal M} \otimes {\cal M} \quad\quad\quad  J^{(2)}_\omega(p,q)=(p', \bar{q}),
\ee
with  
\be
p'_\mu = J_\mu(\omega, p)\,, \quad\quad  
\bar{q}_\mu = J^{(2)}_\mu(\omega, p, q),
\label{J2}
\ee
where $J^{(2)}_\mu(\omega, p, q)$ is determined by the condition 
\be
(p\oplus q)' = p'\oplus \bar{q}.
\label{RP2}
\ee

\end{itemize}

The deformed relativistic kinematics is defined by the (deformed) energy--momentum conservation law of the total momentum and a deformed dispersion relation $C(k)=m^2$ for a particle with momentum $k$ and mass $m$ such that
\be
C(k')=C(k).
\label{RP1}
\ee

When one considers infinitesimal Lorentz transformations
\be
J_\mu(\epsilon, k) = k_\mu + \epsilon_{\alpha\beta} {\cal J}^{\alpha\beta}_\mu(k) \,, \quad\quad
J^{(2)}_\mu(\epsilon, p, q) = q_\mu + \epsilon_{\alpha\beta} {\cal J}^{\alpha\beta}_\mu(p,q)\,,
\label{J2epsilon}
\ee 
the condition (\ref{RP2}) that guarantees the relativistic invariance of the conservation law of momenta becomes
\be
{\cal J}^{\alpha\beta}_\mu(p\oplus q) = \frac{\partial(p\oplus q)_\mu}{\partial p_\nu} \,{\cal J}^{\alpha\beta}_\nu(p) + \frac{\partial(p\oplus q)_\mu}{\partial q_\nu} \,{\cal J}^{\alpha\beta}_\nu(p,q)\,, 
\label{RP-CL}
\ee
which in fact implies 
\be
{\cal J}^{\alpha\beta}_\nu(q) = {\cal J}^{\alpha\beta}_\nu(0,q).
\label{eq:limpto0}
\ee

This result could have been derived directly by putting $p=0$ (and then $p'=0$) in Equation~(5). One has then $\bar{q}=q'$, which is equivalent to Equation~(9). 

Together with the condition (\ref{RP-CL}), we have the condition
\be
\frac{\partial C(p)}{\partial p_\mu} {\cal J}^{\alpha\beta}_\mu(p) = 0\,, \quad
\frac{\partial C(q)}{\partial q_\mu} {\cal J}^{\alpha\beta}_\mu(p,q) = 0\,,
\label{RP-DR}
\ee
from the relativistic invariance of the dispersion relations.

When the total momentum of the two-particle system is $q\oplus p$ instead of $p\oplus q$, all one has to do is exchange the momenta $p\leftrightarrow q$ in all the previous relations.

One could have considered another representation ($\hat{J}^{(2)}$), instead of ($J^{(2)}$), of the Lorentz group in the two particle system with a total momentum ${\cal P}=p\oplus q$,
\be
\hat{J}^{(2)}_\omega: \quad\quad {\cal M} \otimes {\cal M} \to {\cal M} \otimes {\cal M} \quad\quad\quad  \hat{J}^{(2)}_\omega(p,q)=(\bar{p}, q')
\ee
with 
\be
\bar{p}_\mu = \hat{J}^{(2)}_\mu(\omega, p, q) \,, \quad\quad  
q'_\mu = \hat{J}_\mu(\omega, q)\,.
\label{altJ2}
\ee

The condition of relativistic invariance of the conservation law, 
\be
(p\oplus q)' = \bar{p}\oplus q',
\label{RP2-2}
\ee
would be in this case 
\be
\hat{{\cal J}}^{\alpha\beta}_\mu(p\oplus q) = \frac{\partial(p\oplus q)_\mu}{\partial p_\nu} \,\hat{{\cal J}}^{\alpha\beta}_\nu(p,q) + \frac{\partial(p\oplus q)_\mu}{\partial q_\nu} \, \hat{{\cal J}}^{\alpha\beta}_\nu(q)\,, 
\label{RP-CL-bar}
\ee
instead of (\ref{RP-CL}), 
\be
 \hat{J}_\mu(\omega, p) =\hat{J}^{(2)}_\mu(\omega, p, 0)
\ee
instead of \eqref{eq:limpto0},
and the condition for the relativistic invariance of the dispersion relations
\be
\frac{\partial C(p)}{\partial p_\mu} \hat{{\cal J}}^{\alpha\beta}_\mu(p,q) = 0\,, \quad \frac{\partial C(q)}{\partial q_\mu} \hat{{\cal J}}^{\alpha\beta}_\mu(q) = 0\,,
\label{RP-DR-bar}
\ee
instead of (\ref{RP-DR}). 

We see then that there are two different ways to identify a representation of the Lorentz group of transformations in the two-particle system for a given kinematics (i.e., for a given composition law).
In fact, one might consider a more general representation of the Lorentz transformations in the two-particle system, where the Lorentz transformation of the two momenta $p$ and $q$ depend on both momenta. The possibility to have different representations of the Lorentz transformations in the two-particle system was already shown in previous works~\cite{Carmona:2012un,Carmona:2016obd}, where one considers an  expansion in powers of the inverse of a new energy scale, which is a necessary ingredient in a deformation of the kinematics as one can see by purely dimensional arguments.

\section{Derivation of an RDK from the Momentum Space Geometry}
\label{sec:geometry}

A very simple way to derive an RDK is to consider a maximally symmetric momentum space. 
In this geometric approach, the states of a particle defined by its energy and momenta are identified with a subset of points of a manifold. The four coordinates of each point in this subset are just the energy and the momentum of the particle. We have an origin of coordinates, which is in correspondence with the state of a massless particle when its momentum tends to zero. 
Physical states, then, are coordinate dependent, so that different bases in momentum space could in principle represent different physics. There is an ongoing debate in the literature on DSR theories about the meaning of this fact and whether there should be a ``physical basis'' in which physical states should be described~\cite{KowalskiGlikman:2002we}. This discussion goes beyond the scope of the present paper, but we note that the construction we are going to show relies on the definition of isometries, which can be formulated in a coordinate-independent way.

The ten-dimensional group of isometries of the maximally symmetric momentum space geometry can be put in correspondence with the transformations defining a deformed relativistic kinematics. One has transformations ($J_\omega$) which leave one point (origin $k=0$) invariant and translations $T_a$ as isometries
\beq
&& T_a: \:\: {\cal M} \to {\cal M} \quad\quad k'_\mu=[T_a(k)]_\mu\doteq T_\mu(a, k)\,, \nonumber \\
&& J_\omega: \:\: {\cal M} \to {\cal M} \quad\quad 
k'_\mu=[J_\omega(k)]_\mu\doteq J_\mu(\omega, k)\,.
\label{eq:isometries}
\eeq

The isometries $J_\omega$ can be directly identified with the Lorentz transformations acting in momentum space in an RDK. The other isometries (translations, $T_a$) can be used to define the composition law of momenta ($\oplus$) of the RDK. One has two simple ways to define a composition law from a translation,
\be
\text{either } \quad p\oplus q = T_p(q)\,, \quad\quad\text{or } \quad q\oplus p = T_p(q)\,.
\label{T-CL}
\ee

The two options in Equation~\eqref{T-CL} correspond to two different composition laws which are related by the exchange of the two momenta. This ambiguity in the definition of a composition law from a translation can be related with the fact (seen in Section \ref{sec:DRI}) that there are two representations of Lorentz transformations for a given composition law, as we will see at the end of Section~\ref{sec:diagram}. We will remove this ambiguity by taking the first of the two options.

The isometries can be obtained from the inverse of the momentum space metric, $g_{\mu\nu}$, using the set of equations
\be
g_{\mu\nu}(T_a(k)) \,=\, \frac{\partial T_\mu(a, k)}{\partial k_\rho} \frac{\partial T_\nu(a, k)}{\partial k_\sigma} g_{\rho\sigma}(k), \quad\quad
g_{\mu\nu}(J_\omega(k)) \,=\, \frac{\partial J_\mu(\omega, k)}{\partial k_\rho} \frac{\partial J_\nu(\omega, k)}{\partial k_\sigma} g_{\rho\sigma}(k),
\label{T,J}
\ee
that have to be satisfied for any choice of the parameters $a$, $\omega$. 

\medskip

Taking the limit $k\to 0$ in the set of Equation (\ref{T,J}), we get
\be
g_{\mu\nu}(a) \,=\, \left[\lim_{k\to 0} \frac{\partial T_\mu(a, k)}{\partial k_\rho}\right] \, 
\left[\lim_{k\to 0} \frac{\partial T_\nu(a, k)}{\partial k_\sigma}\right] \,\eta_{\rho\sigma}, \quad\quad\quad
\eta_{\mu\nu} \,=\, \left[\lim_{k\to 0} \frac{\partial J_\mu(\omega, k)}{\partial k_\rho}\right] \,  
\left[\lim_{k\to 0} \frac{\partial J_\nu(\omega, k)}{\partial k_\sigma}\right] \,\eta_{\rho\sigma},
\ee
where we considered a system of coordinates such that $g_{\mu\nu}(0)=\eta_{\mu\nu}.$
Now, we can make the identifications 
\be
\lim_{k\to 0} \frac{\partial T_\mu(a, k)}{\partial k_\rho} \,=\, e^\rho_\mu(a), \quad\quad\quad
\lim_{k\to 0} \frac{\partial J_\mu(\omega, k)}{\partial k_\rho} \,=\, L_\mu^\rho(\omega),
\label{e,L}
\ee
where $e_\mu^\rho(k)$ is the (inverse of the) tetrad in momentum space, and $L_\mu^\rho(\omega)$ is the standard $(4\times 4)$ matrix representing the Lorentz transformations with parameters $\omega$.

With the option
\be
p\oplus q \equiv T_p(q),
\label{geom-mcl}
\ee
we obtain from Equation~\eqref{e,L}:
\be
e^\rho_\mu(a) \,=\, \lim_{k\to 0} \frac{\partial(a\oplus k)_\mu}{\partial k_\rho},
\label{magicformula}
\ee
which is a fundamental relationship between (a limit of the derivative of) the composition law and the tetrad in momentum space. 

Considering infinitesimal transformations
\be
T_\mu(\epsilon, k) = k_\mu + \epsilon_\alpha {\cal T}_\mu^\alpha(k), \quad\quad\quad
J_\mu(\epsilon, k) = k_\mu + \epsilon_{\beta\gamma} {\cal J}^{\beta\gamma}_\mu(k),
\ee
where
\be
{\cal T}_\mu^\alpha(k) = \lim_{\epsilon\to 0} \frac{\partial T_\mu(\epsilon,k)}{\partial \epsilon_\alpha} = \lim_{l\to 0} \frac{\partial(l\oplus k)_\mu}{\partial l_\alpha},
\label{Tcal}
\ee
we can use Equation~\eqref{T,J} to get
\be
\frac{\partial g_{\mu\nu}(k)}{\partial k_\rho} {\cal T}^\alpha_\rho(k) \,=\, \frac{\partial{\cal T}^\alpha_\mu(k)}{\partial k_\rho} g_{\rho\nu}(k) +
\frac{\partial{\cal T}^\alpha_\nu(k)}{\partial k_\rho} g_{\mu\rho}(k),
\label{cal(T)}
\ee
\be
\frac{\partial g_{\mu\nu}(k)}{\partial k_\rho} {\cal J}^{\beta\gamma}_\rho(k) \,=\,
\frac{\partial{\cal J}^{\beta\gamma}_\mu(k)}{\partial k_\rho} g_{\rho\nu}(k) +
\frac{\partial{\cal J}^{\beta\gamma}_\nu(k)}{\partial k_\rho} g_{\mu\rho}(k),
\label{cal(J)}
\ee
which is a system of equations for the Killing vectors ${\cal T}^\alpha$, ${\cal J}^{\beta\gamma}$.

Note that if ${\cal T}^\alpha$, ${\cal J}^{\beta\gamma}$ are a solution of the Killing Equations (\ref{cal(T)})--(\ref{cal(J)}), then ${\cal T}'^\alpha = {\cal T}^\alpha + c^\alpha_{\beta\gamma} {\cal J}^{\beta\gamma}$ is also a solution of Equation~\eqref{cal(T)} for any choice of constants $c^\alpha_{\beta\gamma}$, and one has $T'_\mu(\epsilon, 0)=T_\mu(\epsilon, 0)=\epsilon_\mu$. Then, there is an ambiguity in the identification of translations, and also in the deformed composition law. 

It is convenient to introduce coordinates $x^\mu$, canonically conjugated to the momenta, $\{k_\mu,x^\nu\}=\delta^\nu_\mu$, which are, as in SR, the generators of transformations with parameter $a_{\mu}$ in momentum space, $p_\mu \to p_\mu + a_\mu$.
Writing the generators of isometries as 
\be
T^\alpha \,=\, x^\mu {\cal T}^\alpha_\mu(k), \quad\quad\quad J^{\alpha\beta} \,=\, x^\mu {\cal J}^{\alpha\beta}_\mu(k),
\label{T}
\ee
we have
\begin{align}
  &\{T^\alpha, T^\beta\} \,=\, x^\rho \left(\frac{\partial{\cal T}^\alpha_\rho(k)}{\partial k_\sigma} {\cal T}^\beta_\sigma(k) - \frac{\partial{\cal T}^\beta_\rho(k)}{\partial k_\sigma} {\cal T}^\alpha_\sigma(k)\right), \nonumber \\
  &\{T^\alpha, J^{\beta\gamma}\} \,=\, x^\rho \left(\frac{\partial{\cal T}^\alpha_\rho(k)}{\partial k_\sigma} {\cal J}^{\beta\gamma}_\sigma(k) - \frac{\partial{\cal J}^{\beta\gamma}_\rho(k)}{\partial k_\sigma} {\cal T}^\alpha_\sigma(k)\right).
  \label{gen-comm}
\end{align}

Note that the introduction of the coordinates $x^\mu$ are, at this point, merely a mathematical tool that helps one to write the commutators of the generators of isometries in terms of Poisson brackets, as in Equation~\eqref{gen-comm}. We could equally have written
\be
T^\alpha \,=\, \mathcal{T}^\alpha_\mu(k)\frac{\partial}{\partial k_\mu}, \quad\quad\quad J^{\alpha\beta} \,=\, \mathcal{J}^{\alpha\beta}_\mu(k)\frac{\partial}{\partial k_\mu}.
\label{Tb}
\ee

One can check that both definitions, Equations~\eqref{T} and~\eqref{Tb}, are invariant under a momentum change of coordinates when the $x^\mu$ are, as they were introduced above, canonical variables of the momenta. However, the previous notation will turn out to be useful to compare the geometric and the locality approaches to a relativistic deformed kinematics, since the discussion on the locality of interactions in Section~\ref{sec:locality} will take as a starting point an action written in terms of momenta and their canonically conjugated~variables.

The generators $T^\alpha$, $J^{\beta\gamma}$ must close a Lie algebra due to the fact that the isometries are a Lie group of transformations. As we can see from the algebraic perspective, the above mentioned ambiguity in the  identification of translations is just the ambiguity in the choice of basis of $T^\alpha$ in the Lie algebra. Different choices of generators of translations $T^\alpha$  will lead to different deformed composition laws, and then to different relativistic deformed~kinematics.  

Note that Equation~\eqref{T} leads to a transformation of the coordinates $x^\mu$ which turns out to depend on the momentum $k$. However, as we will show later, the loss of locality of the interaction in these coordinates will leads us to indentify a different set of (physical) space-time coordinates.   

\subsection{Construction of a Relativistic Kinematics at the Two-Particle Level}
\label{sec:diagram}

The construction of an RDK from the momentum space geometry given at the one-particle level has to be completed at the two-particle level. We are going to show now that the previous proposal is compatible with the relativity principle if one defines properly the Lorentz transformation in the two-particle system. The sketch of the proof is represented in the following diagram:
\begin{center}
\begin{tikzpicture}
\node (v1) at (-2,1) {$q$};
\node (v4) at (2.7,1) {$\bar q$};
\node (v2) at (-2,-1) {$T_p(q)$};
\node (v3) at (2,-1) {$[T_p(q)]'=T_{p^\prime}(\bar q)$};
\node (v5) at (2.7,-0.8) {};
\draw [->] (v1) edge (v2);
\draw [->] (v4) edge (v5);
\draw [->] (v2) edge (v3);
\node at (-2.4,0) {$T_p$};
\node at (3.1,0) {$T_{p^\prime}$};
\node at (-0.4,-1.4) {$J_\omega$};
\end{tikzpicture}
\end{center}
where $J_\omega$, $T_p$, $T_{p'}$ are different isometries, and $p'=J_\omega(p)$. The point $\bar{q}$ is defined from the~condition
\be
J_\omega(T_p(q)) \,=\, T_{p^\prime}(\bar{q}).
\label{qbar1}
\ee

When $q=0$, one has $\bar{q}=0$, and when $p=0$, $\bar{q}=q'$. The case $q\neq 0$, $p\neq 0$ leads to a point $\bar{q}$ which is obtained from $q$ by a composition of three isometries (the translation $T_p$, a Lorentz transformation $J_\omega$, and the inverse of the translation $T_{p'}$); therefore, it is obtained from the original point $q$ by applying an isometry. This means that $q$ and $\bar{q}$ are at the same distance from the origin and, if one defines $C(k)$ as the square of the distance from the origin to the point $k$, then
\be
C(q) \,=\, C(\bar{q}).
\label{qbar2}
\ee

Since the isometry $q\to\bar{q}$ has the property of leaving the origin invariant, we can identify the momenta $(p', \bar{q})$ as the Lorentz transformed momenta of $(p, q)$, and then \linebreak Equations~\eqref{qbar1} and \eqref{qbar2} tell us that the deformed kinematics defined by $C$ and $\oplus$ is a relativistic deformed kinematics. Indeed, Equation~(\ref{qbar1}) implies the invariance of a conservation law for the total momentum of the two-particle system, that is identified as
\be
\mathcal{P}=T_p(q),
\label{totalP}
\ee
and Equation~(\ref{qbar2}), together with $C(p)=C(p')$, shows that the dispersion relation of the particles is also Lorentz invariant. 

The definition of $\bar{q}$ in Equation~(\ref{qbar1}) also implies that the Lorentz transformation of two momenta depends on both momenta in a nontrivial way, which is determined by a composition of isometries (translations and the Lorentz transformation of one momentum). In fact the solution of Equation~\eqref{qbar1} for $\bar{q}$ as a function of $\omega$, $p$ and $q$, when we consider infinitesimal parameters $\omega=\epsilon$,   
\be
\bar{q}_\mu = q_\mu + \epsilon_{\alpha\beta} {\cal J}_\mu^{\alpha\beta}(p, q) + {\cal O}(\epsilon^2)\,,
\ee
allows one to determine the functions ${\cal J}_\mu^{\alpha\beta}(p, q)$ which give the representation of the Lorentz transformations in the two-particle system for an RDK. 

Note that from the identification of the composition law as in Equation~\eqref{geom-mcl} and the relation between translations and total momentum Equation~\eqref{totalP}, we see that the Lorentz transformation in the two-particle system defined above corresponds to the representation~\eqref{J2} and not to the representation~\eqref{altJ2}. However, one could have considered an alternate metric $\hat{g}$, with translations $\hat T$ such that $\hat{T}_q(p)=T_p(q)$.
Since $\hat T$ and $T$ are related by an exchange of the two momenta $p$ and $q$, the ambiguity of taking $p\oplus q=T_p(q)$ or $p\oplus q=\hat{T}_p(q)$ as a way of associating a translation with a given composition law is equivalent to the ambiguity we had in Equation~\eqref{T-CL} in associating a composition law to a translation. We can again remove this ambiguity by taking the first option, $p\oplus q=T_p(q)=\hat{T}_q(p)$. However, then, one can use the isometries $\hat J$ of $\hat{g}$ that leave the origin invariant to define the one-particle Lorentz transformations, and the two-particle Lorentz transformations would be given by the analogous diagram considered above but with $\hat{T}_q(p)$ and $\hat{J}_\omega$ while the total momentum would still be $p\oplus q$. 
The Lorentz transformation of the two momenta, $(p,q)\to (\bar{p},q')$, would then be given by the second representation of the Lorentz group considered in Section~\ref{sec:DRI}, Equation~\eqref{altJ2}.

\subsection{Isotropic Relativistic Deformed Kinematics: The \texorpdfstring{$\kappa$}{k}-Poincaré Example}
\label{sec:examples}

In this subsection, we explain how different representations in the algebra of translations (different choices of bases) lead to different kinematics (different composition laws). The general form of the algebra of translations in the case of an isotropic relativistic deformed kinematics contains two parameters:
\be
\{T^0, T^i\} \,=\, \frac{c_1}{\Lambda} T^i + \frac{c_2}{\Lambda^2} J^{0i}, \quad\quad\quad \{T^i, T^j\} \,=\, \frac{c_2}{\Lambda^2} J^{ij},
\label{isoRDK}
\ee
where we assume that the generators of isometries leaving the origin invariant $J^{\alpha\beta}$ have been chosen to satisfy the standard Lorentz algebra (the Poisson brackets of $T^\alpha$ and $J^{\beta\gamma}$ are then fixed by Jacobi identities). For each choice of this algebra (i.e., for each choice of $(c_1/\Lambda)$ and $(c_2/\Lambda^2)$) and for each choice of an isotropic metric, one has to find the isometries of the metric in which the generators satisfy the chosen algebra. These isometries define an isotropic relativistic deformed kinematics. 

As it was explained in~\cite{Carmona:2019fwf}, the choice $c_1=0$ leads to the Snyder kinematics~\cite{Battisti:2010sr} while, when none of the $c_1$, $c_2$ parameters are null, one obtains the kinematics of the hybrid models~\cite{Meljanac:2009ej}. 
As we now show, the derivation of an RDK from the momentum space geometry is particularly simple if the corresponding composition law is associative; in this case, translations form a subgroup within the group of isometries, so that $c_2=0$ and then    
\be
\{T^0, T^i\} \,=\, \frac{1}{\Lambda} T^i,
\label{Talgebra}
\ee
where we reabsorbed the coefficient $c_1$ into a redefinition of the scale $\Lambda$. The choice of a positive coefficient is due to the fact that we want to make the correspondence with $\kappa$-Poincaré kinematics explicit, as we will see. 

Together with the translations $k \to T_a(k) = (a\oplus k)$, one can consider the transformations with generators $\bar{T}^\alpha$ defined from the tetrad in momentum space by
\be
\bar{T}^\alpha \,=\, x^\mu e^\alpha_\mu(k).
\label{Tbar}
\ee

For the infinitesimal transformation with parameter $\epsilon$, one has \be
k^\prime_\mu = k_\mu+\epsilon_\alpha \{k_\mu,\bar{T}^\alpha\}=k_\mu+ \epsilon_\alpha e^\alpha_\mu(k) = 
k_\mu + \epsilon_\alpha \lim_{q\to 0} \frac{\partial(k\oplus q)_\mu}{\partial q_\alpha} = (k\oplus \epsilon)_\mu.
\ee

The finite transformation $\bar{T}_a$ with generators $\bar{T}^\alpha$ is then 
\be
\bar{T}_a(k) = k\oplus a\,.
\label{Tbar-finite}
\ee

A result of differential geometry~\cite{Chern:1999jn} is that, when the generators of translations $T^\alpha$ satisfy the Lie algebra (\ref{Talgebra}), then the $\bar{T}^\alpha$ are also the generators of a Lie algebra, 
\be
 \{\bar{T}^0, \bar{T}^i\} \,=\, - \frac{1}{\Lambda} \bar{T}^i ,
\label{Tbaralgebra}
\ee   
which is the Lie algebra of the generators of translations up to a sign. Both algebras \eqref{Talgebra} and \eqref{Tbaralgebra} are simply the algebra for the coordinates of $\kappa$-Minkowski space-time.  Then, in order to determine a tetrad $e^\alpha_\mu(k)$ compatible with the algebra (\ref{Tbaralgebra}), we have to find a representation of $\kappa$-Minkowski space-time coordinates in terms of canonical phase space coordinates. \linebreak A~particularly simple choice (that, as we will see, corresponds to the so-called bicrossproduct basis of $\kappa$-Poincaré algebra) is
\be
e^0_0(k) \,=\, 1, \quad\quad\quad e^0_i(k) \,=\, e^i_0(k) \,=\, 0, \quad\quad\quad e^i_j (k) \,=\, \delta^i_j e^{- k_0/\Lambda}.
\label{bicross-tetrad}
\ee

\textls[-20]{In order to obtain the finite translations $T_\mu(a,k)$ (with generators satisfying \mbox{Equation~\eqref{Talgebra}})}, one can try to generalize the first equation in Equation~\eqref{e,L} to define a transformation that does not change the form of the tetrad:
\be
e_\mu^\alpha(T(a, k)) \,=\, \frac{\partial T_\mu(a, k)}{\partial k_\nu} \,e_\nu^\alpha(k).
\label{T(a,k)}
\ee

Then, if $T_\mu(a,k)$ is a solution to this equation, it will be an isometry (since it leaves the tetrad invariant and then the metric) and translations will form a group because the composition of two transformations leaving the tetrad invariant has the same property. It is easy to see that Equation~\eqref{T(a,k)} can be solved, allowing us to determine the finite translations:
\be
T_0(a, k) \,=\, a_0 + k_0, \quad\quad\quad T_i(a, k) \,=\, a_i + k_i e^{- a_0/\Lambda},
\ee
leading to the composition law of momenta 
\be
(p\oplus q)_0 \,=\, T_0(p, q) \,=\, p_0 + q_0, \quad\quad\quad
(p\oplus q)_i \,=\, T_i(p, q) \,=\, p_i + q_i e^{- p_0/\Lambda},
\label{kappa-dcl}
\ee
which is just the DCL of $\kappa$-Poincaré kinematics in the bicrossproduct basis.

The dispersion relation can be obtained by asking the function $C(k)$ to be invariant under Lorentz transformations:
\be
\frac{\partial C(k)}{\partial k_\mu} \,{\cal J}^{\alpha\beta}_\mu(k) \,=\, 0 ,
\ee
where ${\cal J}^{\alpha\beta}_\mu$ are the Killing vectors satisfying Equation~\eqref{cal(J)} with the metric $g_{\mu\nu}(k)=e^\alpha_\mu(k)\eta_{\alpha\beta}e^\beta_\nu(k)$ defined by the tetrad~\eqref{bicross-tetrad}:
\be
0 \,=\, \frac{{\cal J}^{\alpha\beta}_0(k)}{\partial k_0}, \quad\quad
0 \,=\, - \frac{{\cal J}^{\alpha\beta}_0(k)}{\partial k_i} e^{- 2k_0/\Lambda} + \frac{{\cal J}^{\alpha\beta}_i(k)}{\partial k_0}, \quad\quad
+ \frac{2}{\Lambda} {\cal J}^{\alpha\beta}_0(k) \delta_{ij} \,=\, - \frac{\partial{\cal J}^{\alpha\beta}_i(k)}{\partial k_j} - \frac{\partial{\cal J}^{\alpha\beta}_j(k)}{\partial k_i} .
\ee
The solution for ${\cal J}^{\alpha\beta}_\mu(k)$ is
 \be
{\cal J}^{0i}_0(k) \,=\, -k_i, \quad \quad \quad {\cal J}^{0i}_j(k)\,=\,  \delta^i_j \,\frac{\Lambda}{2} \left[e^{- 2 k_0/\Lambda} - 1 - \frac{\vec{k}^2}{\Lambda^2}\right] + \,\frac{k_i k_j}{\Lambda},
\ee 
and then 
\be
C(k) \,=\, \Lambda^2 \left(e^{k_0/\Lambda} + e^{-k_0/\Lambda} - 2\right) - e^{ k_0/\Lambda} \vec{k}^2  \,,
\label{kappa-C}
\ee
which is the same dispersion relation appearing in  $\kappa$-Poincaré kinematics in the bicrossproduct basis.

The last ingredient we need in order to define the kinematics is the Lorentz transformation for two momenta. Since the first momentum transformation does not depend on the other, from Equation~\eqref{RP2},
it is easy to determine $\bar{q}$ by equating the terms linear in $\epsilon_{\alpha\beta}$ on both sides of this equation. One finds
\be
\epsilon_{\alpha\beta} {\cal J}^{\alpha\beta}_\mu(p\oplus q) \,=\, \epsilon_{\alpha\beta} \frac{\partial(p\oplus q)_\mu}{\partial p_\nu} {\cal J}^{\alpha\beta}_\nu(p) + \frac{\partial(p\oplus q)_\mu}{\partial q_\nu} (\bar{q}_\nu - q_\nu).
\ee

From the composition law (\ref{kappa-dcl}) with the minus sign in the exponent, we have
\begin{align}
& \frac{\partial(p\oplus q)_0}{\partial p_0} \,=\, 1, \quad\quad
\frac{\partial(p\oplus q)_0}{\partial p_i} \,=\, 0, \quad\quad
\frac{\partial(p\oplus q)_i}{\partial p_0} \,=\, - \frac{q_i}{\Lambda} e^{-p_0/\Lambda}, \quad\quad
\frac{\partial(p\oplus q)_i}{\partial p_j} \,=\, \delta_i^j, \\
& \frac{\partial(p\oplus q)_0}{\partial q_0} \,=\, 1, \quad\quad  \frac{\partial(p\oplus q)_0}{\partial q_i} \,=\, 0, \quad\quad
\frac{\partial(p\oplus q)_i}{\partial q_0} \,=\, 0, \quad\quad  \frac{\partial(p\oplus q)_i}{\partial q_j} \,=\, \delta_i^j e^{-p_0/\Lambda}.
\end{align}
Then we find
\begin{align}
& \bar{q}_0 \,=\, q_0 + \epsilon_{\alpha\beta} \left[{\cal J}^{\alpha\beta}_0(p\oplus q) - {\cal J}^{\alpha\beta}_0(p)\right], \nonumber \\
& \bar{q}_i \,=\, q_i + \epsilon_{\alpha\beta} \, e^{p_0/\Lambda} \, \left[{\cal J}^{\alpha\beta}_i(p\oplus q) - {\cal J}^{\alpha\beta}_i(p) + \frac{q_i}{\Lambda} e^{-p_0/\Lambda} {\cal J}^{\alpha\beta}_0(p)\right],
\end{align}
and one can check that this is the same Lorentz transformation of two momenta $(p, q) \to (p', \bar{q})$ obtained from the coproduct of the Lorentz generators of  $\kappa$-Poincaré Hopf algebra in the bicrossproduct basis. 

We have seen then that $\kappa$-Poincaré kinematics in the bicrossproduct basis is precisely the deformed kinematics obtained from a de Sitter momentum space when one uses coordinates such that the tetrad takes the form in (\ref{bicross-tetrad}) and one defines the translations as the isometries that leave the tetrad invariant. Different choices of tetrad, corresponding to different representations of $\kappa$-Minkowski space-time coordinates in terms of canonical phase-space variables, will lead to $\kappa$-Poincaré kinematics in different bases, that is, they will correspond to different choices of coordinates in momentum space of the deformed kinematics defined by the algebra~\eqref{Talgebra}.

We can also check that the momentum metric derived from tetrad (\ref{bicross-tetrad}) is
\be
g_{00}(k) \,=\, 1, \quad\quad\quad g_{0i}(k) \,=\, g_{i0}(k) \,=\, 0, \quad\quad\quad g_{ij}(k) \,=\, - \delta_{ij} e^{- 2k_0/\Lambda},
\label{bicross-metric}
\ee
which is the metric in the comoving coordinates system of a de Sitter space~\cite{Gubitosi:2013rna} with curvature $(12/\Lambda^2)$.

With all this, we can see that, for the case of an associative composition law, the geometric interpretation of the deformed kinematics allows us to obtain the same results as those corresponding to the $\kappa$-Poincaré Hopf algebra in the bicrossproduct basis~\cite{KowalskiGlikman:2002we}.

\section{Generalized Space-Time and Locality of Interactions with an RDK}
\label{sec:locality}

We now review the relationship between a deformed kinematics and the property of locality of interactions. We start by considering the relative locality action proposed in~\cite{AmelinoCamelia:2011bm}. The classical action we show here depicts the free propagation of two particles, described by the deformed dispersion relation; its interaction, controlled by the deformed composition law; and the free propagation after the collision, again characterized by the dispersion relation:
\begin{align}
S^{(2)} \,=&\, \int_{-\infty}^0 d\tau \sum_{i=1,2} \left[x_{-(i)}^\mu(\tau) \dot{p}_\mu^{-(i)}(\tau) + N_{-(i)}(\tau) \left[C(p^{-(i)}(\tau)) - m_{-(i)}^2\right]\right] \nonumber \\   
& + \int^{\infty}_0 d\tau \sum_{j=1,2} \left[x_{+(j)}^\mu(\tau) \dot{p}_\mu^{+(j)}(\tau) + N_{+(j)}(\tau) \left[C(p^{+(j)}(\tau)) - m_{+(j)}^2\right]\right] \nonumber \\
& + \xi^\mu \left[{\cal P}^+_\mu(0) - {\cal P}^-_\mu(0)\right],
\label{S2}
\end{align}
where $\dot{a}\doteq (da/d\tau)$ is a derivative of the variable $a$ with respect to the parameter $\tau$ along the trajectory of the particle, $x_{-(i)}$ ($x_{+(j)}$) is the space-time coordinates of the in-state (out-state) particles, $p^{-(i)}$ ($p^{+(j)}$) is their four-momenta, $m_{-(i)}$ ($m_{+(j)}$) is their masses, ${\cal P}^-$ (${\cal P}^+$) is the total four-momentum of the in-state (out-state) defining the DCL, $C(k)$ is the function of a four-momentum $k$ defining the DDR, $\xi^\mu$ is Lagrange multipliers that implement the energy-momentum conservation in the interaction, and $N_{-(i)}$ ($N_{+(j)}$) is Lagrange multipliers implementing the dispersion relation of in-state (out-state) particles.

Applying the variational principle to the action (\ref{S2}), one finds that all the momenta are independent of $\tau$ and obtains the end (starting) space-time coordinates of the trajectories of the in-state (out-state) particles
\be
x_{-(i)}^\mu(0) \,=\, \xi^\nu \frac{\partial {\cal P}^-_\nu}{\partial p^{-(i)}_\mu}, \quad\quad\quad
x_{+(j)}^\mu(0) \,=\, \xi^\nu \frac{\partial {\cal P}^+_\nu}{\partial p^{+(j)}_\mu}.
\ee

In the case where the composition law is not deformed, i.e., the SR limit, one has $x_{-(i)}^\mu(0)=x_{+(j)}^\mu(0)=\xi^\mu$, so that the worldlines of the four particles cross at the same space-time point with coordinates $\xi^\mu$, leading to  local interactions. Considering a deformed composition law leads then to the loss the locality of interactions. 

\subsection{Generalized Space-Time Coordinates}

We now wonder whether there are new space-time coordinates in which interactions are local~\cite{Carmona:2019vsh}. Considering the coordinates of the two particles, either in the in-state or out-state, and then omitting the index $-$,$+$, let us define new space-time coordinates
\be
\tilde{x}^\alpha_{(1)} \,=\, x^\mu_{(1)} \varphi^\alpha_\mu(p^{(1)}) + x^\mu_{(2)} \varphi^{(2)\alpha}_{(1)\mu}(p^{(2)}), \quad\quad
\tilde{x}^\alpha_{(2)} \,=\, x^\mu_{(2)} \varphi^\alpha_\mu(p^{(2)}) + x^\mu_{(1)} \varphi^{(1)\alpha}_{(2)\mu}(p^{(1)}),
\label{generalized-spacetime}
\ee
and ask that they satisfy  $\tilde{x}^\alpha_{(1)}(0)=\tilde{x}^\alpha_{(2)}(0)$, having then local interactions in the generalized space-time with coordinates $\tilde{x}^\alpha$. We assume that $\varphi^{(2)\alpha}_{(1)\mu}(0) = \varphi^{(1)\alpha}_{(2)\mu}(0) = 0$ so that the system of two particles reduces, when one of the two momenta is zero, to one particle with new space-time coordinates $\tilde{x}^\alpha = x^\mu \varphi^\alpha_\mu(k)$. One also assumes that $\varphi^\alpha_\mu(0)=\delta^\alpha_\mu$ so that the new space-time coordinates coincide with the coordinates $x$ in the limit $k\to 0$.  

When the total momentum of the two-particle system is
\be
{\cal P}_\mu \,=\, (p^{(1)}\oplus p^{(2)})_\mu,
\label{totalmomentum}
\ee
the conditions of having local interactions (crossing of in- and out-state worldlines in a point) require that the functions $\varphi^\alpha_\mu(k)$, $\varphi^{(2)\alpha}_{(1)\mu}(k)$ and $\varphi^{(1)\alpha}_{(2)\mu}(k)$ satisfy the set of equations
\be
\begin{split}
    &\frac{\partial(p^{(1)}\oplus p^{(2)})_\mu}{\partial p^{(1)}_\nu} \,\varphi^\alpha_\nu(p^{(1)}) \,+\, 
\frac{\partial(p^{(1)}\oplus p^{(2)})_\mu}{\partial p^{(2)}_\nu} \,\varphi^{(2)\alpha}_{(1)\nu}(p^{(2)}) \,=\, \\
&\frac{\partial(p^{(1)}\oplus p^{(2)})_\mu}{\partial p^{(2)}_\nu} \,\varphi^\alpha_\nu(p^{(2)}) \,+\, 
\frac{\partial(p^{(1)}\oplus p^{(2)})_\mu}{\partial p^{(1)}_\nu} \,\varphi^{(1)\alpha}_{(2)\nu}(p^{(1)}).
\end{split}
\label{loc-eq}
\ee

Additionally, due to the new conservation law
\be
(p^{-(1)}\oplus p^{-(2)})_\mu \,=\, (p^{+(1)}\oplus p^{+(2)})_\mu,
\ee
the crossing of the worldlines of the four particles at a point requires the left-hand side and the right-hand side of Equation~(\ref{loc-eq}) to depend on the two four-momenta only through the combination $(p^{(1)}\oplus p^{(2)})$. Using the conditions  $\varphi^{(2)\alpha}_{(1)\mu}(0) = \varphi^{(1)\alpha}_{(2)\mu}(0) = 0$, one concludes that, in fact, both sides of Equation~(\ref{loc-eq}) should be equal to $\varphi^\alpha_\mu(p^{(1)}\oplus p^{(2)})$~\cite{Carmona:2019vsh}.

Taking the limit $p^{(1)}\to 0$ or $p^{(2)}\to 0$ in the locality Equation~\eqref{loc-eq}, one has
\be
\varphi^{(2)\alpha}_{(1)\mu}(p^{(2)}) \,=\, \varphi^\alpha_\mu(p^{(2)}) - \lim_{k\to 0}  \frac{\partial(k\oplus p^{(2)})_\mu}{\partial k_\alpha}, \quad\quad
\varphi^{(1)\alpha}_{(2)\mu}(p^{(1)}) \,=\, \varphi^\alpha_\mu(p^{(1)}) - \lim_{k\to 0} \frac{\partial(p^{(1)}\oplus k)_\mu}{\partial k_\alpha}.
\label{eq:phi12}
\ee
Plugging these functions $\varphi^{(2)}_{(1)}$, $\varphi^{(1)}_{(2)}$ into Equation~\eqref{loc-eq}, one finds
\begin{align}
& \frac{\partial(p^{(1)}\oplus p^{(2)})_\mu}{\partial p^{(2)}_\nu} \, \lim_{k\to 0} \frac{\partial(k\oplus p^{(2)})_\nu}{\partial k_\alpha} \,=\, \frac{\partial(p^{(1)}\oplus p^{(2)})_\mu}{\partial p^{(1)}_\nu}  \, \lim_{k\to 0} \frac{\partial(p^{(1)}\oplus k)_\nu}{\partial k_\alpha} \,=\, \nonumber \\ & \:\:\: \frac{\partial(p^{(1)}\oplus p^{(2)})_\mu}{\partial p^{(1)}_\nu} \varphi^\alpha_\nu(p^{(1)}) + 
\frac{\partial(p^{(1)}\oplus p^{(2)})_\mu}{\partial p^{(2)}_\nu} \varphi^\alpha_\nu(p^{(2)}) - \varphi^\alpha_\mu(p^{(1)}\oplus p^{(2)})\,.
\label{loc-oplus-varphi}
\end{align}

The first equality is a set of equations that a composition law must satisfy in order to be able to implement locality, while the second one establishes a relation between the composition law and the functions $\varphi^\alpha_\mu$ that define the new space-time coordinates in the one-particle system.

Considering the identities 
\begin{align}
& \frac{\partial(p^{(1)}\oplus p^{(2)})_\mu}{\partial p^{(2)}_\nu} \, \lim_{k\to 0} \frac{\partial(k\oplus p^{(2)})_\nu}{\partial k_\alpha} \,=\, \lim_{k\to 0}  \frac{\partial(p^{(1)}\oplus (k\oplus p^{(2)}))_\mu}{\partial (k\oplus p^{(2)})_\nu} \, \frac{\partial(k\oplus p^{(2)})_\nu}{\partial k_\alpha} \,=\, \lim_{k\to 0} \frac{\partial(p^{(1)}\oplus (k\oplus p^{(2)}))_\mu}{\partial k_\alpha}, \\ \nonumber
& \frac{\partial(p^{(1)}\oplus p^{(2)})_\mu}{\partial p^{(1)}_\nu}  \, \lim_{k\to 0} \frac{\partial(p^{(1)}\oplus k)_\nu}{\partial k_\alpha} \,=\,  \lim_{k\to 0} \,\frac{\partial((p^{(1)} \oplus k)\oplus p^{(2)})_\mu}{\partial (p^{(1)}\oplus k)_\nu}  \, \frac{\partial(p^{(1)}\oplus k)_\nu}{\partial k_\alpha} \,=\, \lim_{k\to 0} \frac{\partial((p^{(1)}\oplus k)\oplus p^{(2)})_\mu}{\partial k_\alpha},
\end{align} 
the first equality in (\ref{loc-oplus-varphi}) leads to 
\be
\lim_{k\to 0} \frac{\partial(p^{(1)}\oplus (k\oplus p^{(2)}))_\mu}{\partial k_\alpha} \,=\, \lim_{k\to 0} \frac{\partial((p^{(1)}\oplus k)\oplus p^{(2)})_\mu}{\partial k_\alpha},
\ee
which implies
\be
(p^{(1)}\oplus \epsilon)\oplus p^{(2)} \,=\, p^{(1)}\oplus(\epsilon\oplus p^{(2)}),
\label{eq:associativity}
\ee
where $\epsilon$ is an infinitesimal momentum.

Equation~\eqref{loc-oplus-varphi} does not allow us to completely determine the function $\varphi^\alpha_\mu(p)$ from the composition law of momenta. To do so, we can add the condition that $\varphi^{(2)\alpha}_{(1)\mu}(p^{(2)})=0$ in Equation~\eqref{generalized-spacetime} so that the generalized space-time coordinates of particle (1) are simply 
\be
\tilde{x}_{(1)}^\alpha = x_{(1)}^\mu \varphi^\mu_\alpha(p^{(1)})\,.
\ee 
Then, Equation~\eqref{eq:phi12} gives
\begin{equation}
\varphi^\alpha_\mu(p^{(2)}) \,=\, \lim_{k\to 0} \frac{\partial(k\oplus p^{(2)})_\mu}{\partial k_\alpha}\,,
\label{eq:phi-magic-L}
\end{equation}
and then
\begin{equation}
\begin{split}
\frac{\partial(p^{(1)}\oplus p^{(2)})_\mu}{\partial p^{(1)}_\nu} \,\varphi^\alpha_\nu(p^{(1)}) \,=\, \frac{\partial(p^{(1)}\oplus p^{(2)})_\mu}{\partial p^{(1)}_\nu} \,\lim_{k\to 0} \frac{\partial(k\oplus p^{(1)})_\nu}{\partial k_\alpha} \,&=\, \lim_{k\to 0} \left[\frac{\partial((k\oplus p^{(1)})\oplus p^{(2)})_\mu}{\partial(k\oplus p^{(1)})_\nu} \,\frac{\partial(k\oplus p^{(1)})_\nu}{\partial k_\alpha}\right]\\
\,&=\, \lim_{k\to 0} \frac{\partial((k\oplus p^{(1)})\oplus p^{(2)})_\mu}{\partial k_\alpha},  \\
\\
\frac{\partial(p^{(1)}\oplus p^{(2)})_\mu}{\partial p^{(2)}_\nu} \,\varphi^\alpha_\nu(p^{(2)}) \,=\, \frac{\partial(p^{(1)}\oplus p^{(2)})_\mu}{\partial p^{(2)}_\nu} \,\lim_{k\to 0} \frac{\partial(k\oplus p^{(2)})_\nu}{\partial k_\alpha} \,&=\, \lim_{k\to 0} \left[\frac{\partial(p^{(1)}\oplus(k\oplus p^{(2)})_\mu}{\partial(k\oplus p^{(2)})_\nu} \,\frac{\partial(k\oplus p^{(2)})_\nu}{\partial k_\alpha}\right]\\ \,&=\, \lim_{k\to 0} \frac{\partial(p^{(1)}\oplus(k\oplus p^{(2)})_\mu}{\partial k_\alpha}, 
\end{split}
\label{eq:associativity_locality_1}
\end{equation}

\begin{equation}
\varphi^\alpha_\mu(p^{(1)}\oplus p^{(2)}) \,=\, \lim_{k\to 0} \frac{k\oplus\partial((p^{(1)}\oplus p^{(2)}))_\mu}{\partial k_\alpha}\,.
\label{eq:associativity_locality_varphi}
\end{equation}

In this way, the relations of compatibility with locality, Equation~\eqref{loc-oplus-varphi}, can be written as
\be
\begin{split}
&\lim_{k\to 0} \frac{\partial(p^{(1)}\oplus (k\oplus p^{(2)}))_\mu}{\partial k_\alpha} \,=\, \lim_{k\to 0} \frac{\partial((p^{(1)}\oplus k)\oplus p^{(2)})_\mu}{\partial k_\alpha}\,=\, \\ & \lim_{k\to 0} \frac{\partial((k\oplus p^{(1)})\oplus p^{(2)})_\mu}{\partial k_\alpha} \,+\, \lim_{k\to 0} \frac{\partial(p^{(1)}\oplus(k\oplus p^{(2)}))_\mu}{\partial k_\alpha} \,-\,\lim_{k\to 0} \frac{\partial(k\oplus(p^{(1)}\oplus p^{(2)}))_\mu}{\partial k_\alpha}.
\end{split}
\label{eq:associativity_locality_2}
\ee

This shows that any associative composition law is compatible with locality. 


If we make the alternative choice $\varphi^{(1)\alpha}_{(2)\mu}(p^{(1)})=0$ in Equation~\eqref{eq:phi12}, we get
\begin{equation}
\varphi^\alpha_\mu(p^{(1)}) \,=\, \lim_{k\to 0} \frac{\partial(p^{(1)}\oplus k)_\mu}{\partial k_\alpha}\,.
\label{eq:phi-magic-R}
\end{equation}

The arguments leading to the conclusion that, using Equation~\eqref{eq:phi-magic-L},  any associative composition law is compatible with locality can be repeated if one uses Equation~\eqref{eq:phi-magic-R}.

To summarize, the implementation of locality has led us to identify two possible choices for the generalized space-time  coordinates of one particle: 
\begin{equation}
\tilde{x}^\alpha = x^\mu \lim_{l\to 0} \frac{\partial{(l\oplus k)}_\mu}{\partial l_\alpha}, \quad\quad\quad \text{or} \quad\quad\quad
\tilde{x}^\alpha = x^\mu \lim_{l\to 0} \frac{\partial{(k\oplus l)}_\mu}{\partial l_\alpha}.
\label{gensptimechoices}
\end{equation}

This generalized space-time coordinates can be used as generators of infinitesimal transformations in momentum space:
\begin{equation}
\delta k_\mu = \epsilon_\alpha \{k_\mu, \tilde{x}^\alpha\} = (\epsilon\oplus k)_\mu - k_\mu, \quad\quad\quad \text{or} \quad\quad\quad 
\delta k_\mu = \epsilon_\alpha \{k_\mu, \tilde{x}^\alpha\} = (k\oplus \epsilon)_\mu - k_\mu.
\end{equation}

If the composition law is associative, the generalized space-time coordinates are the generators of a group of finite transformations (with parameters $a$) in momentum space:
\begin{equation}
k'_\mu = (a\oplus k), \quad\quad\quad \text{or} \quad\quad\quad
k'_\mu = (k\oplus a).
\end{equation}

Then, we see that the two choices of generalized space-time coordinates implementing locality are in one to one correspondence with the two choices to define a composition law of momenta from a translation in momentum space in the geometric framework.

\subsection{Propagation of a Massless Particle in Space-Time}

One can try to complement the identification of the generalized space-time by considering, together with the locality of interactions, the propagation of a massless particle. In the case of SR, an energy-independent velocity of propagation of a massless particle is a crucial property of the notion of space-time. This leads to the consideration of what happens when one considers the propagation of a particle in the generalized space-time defined by the locality of interactions. 

The model for the propagation of a massless particle can be obtained from \mbox{Equation (\ref{S2})} by considering one particle instead of two particles. Then, one has 
\begin{align}
S^{(1)} =& \int_{-\infty}^0 d\tau \left[x_-^\mu(\tau) \dot{p}^-_\mu(\tau)+N(\tau) C(p^-(\tau))\right] \nonumber \\
&+ \int^{\infty}_0 d\tau \left[x_+^\mu(\tau) \dot{p}^+_\mu(\tau)+N(\tau) C(p^+(\tau))\right] + \xi^\mu \left[p^+_\mu(0)-p^-_\mu(0)\right].
\label{S1}
\end{align}

One has in this case 
\be
x_-^\mu(0)=\xi^\mu = x_+^\mu(0).
\ee

Then, one can introduce 
\begin{align}
& x^\mu(\tau)=x_-^\mu(\tau) \quad \text{for} \, \tau <0 \quad x^\mu(\tau)=x_+^\mu(\tau) \quad \text{for} \, \tau >0, \nonumber \\
& p^\mu(\tau)=p_-^\mu(\tau) \quad \text{for} \, \tau <0 \quad p^\mu(\tau)=p_+^\mu(\tau) \quad \text{for} \, \tau >0,
\end{align}
and the action in (\ref{S1}) can be written in a simplified form:
\be
S_0 = \int_{-\infty}^\infty d\tau \left[x^\mu(\tau) \dot{p}_\mu(\tau) + N(\tau) C(p(\tau)\right]\,.
\label{S0}
\ee

Applying the variational principle to the action (\ref{S0}), one has
\be
\dot{p}_\mu=0, \quad\quad \dot{x}^\mu = N \frac{\partial C(p)}{\partial p_\mu}, \quad\quad C(p)=0,
\ee
and using the generalized space-time coordinates $\tilde{x}^\alpha$, one finds
\be
\dot{\tilde{x}}^\alpha \eta_{\alpha\beta} \dot{\tilde{x}}^\beta = N^2 \left[\frac{\partial C(p)}{\partial p_\mu} \varphi^\alpha_\mu(p) \eta_{\alpha\beta} \varphi^\beta_\nu(p) \frac{\partial C(p)}{\partial p_\nu}\right]\,,
\label{xtilde-square}
\ee
a result derived from a geometrical point of view in Reference~\cite{Relancio:2020zok}.

The generalized space-time coordinates $\tilde{x}^\alpha=x^\mu \varphi^\alpha_\mu(p)$ (and then the expression \mbox{in (\ref{xtilde-square})}) are invariant under a canonical change of phase space variables corresponding to a nonlinear change in momentum variables. Instead of using the dispersion relation (\ref{kappa-C}) and composition law (\ref{kappa-dcl}) of $\kappa$-Poincaré in the bicrossproduct basis, it is easier to use a basis where the deformation of the composition law is proportional to $(1/\Lambda)$
\be
(p\oplus q)_\mu = p_\mu + \left(1-\frac{p_0}{\Lambda}\right) q_\mu, \quad\quad C(p) = \frac{p_0^2 - \vec{p}^2}{1-\frac{p_0}{\Lambda}}.
\label{DRK1}
\ee

This is the basis found directly from the locality relations~\cite{Carmona:2019fwf}.   
The derivatives $(\partial C/\partial p)$ when $C=0$ are
\be
\frac{\partial C}{\partial p_0} = \frac{2 p_0}{1-\frac{p_0}{\Lambda}}, \quad\quad 
\frac{\partial C}{\partial p_i} = - \frac{2 p_i}{1-\frac{p_0}{\Lambda}}.
\ee

For the first choice of space-time  coordinates in Equation~\eqref{gensptimechoices}, one has
\be
\varphi^\alpha_\mu(p) = \lim_{l\to 0} \frac{\partial(l\oplus p)_\mu}{\partial l_\alpha}, \ee
and using the composition law in (\ref{DRK1}), one gets
\be
\varphi^0_0=1-\frac{p_0}{\Lambda},\quad \varphi_0^i=0,\quad \varphi_i^0=-\frac{p_i}{\Lambda},\quad \varphi^j_i=\delta_i^j.
\ee

Equation~(\ref{xtilde-square}) gives in this case
\be
\dot{\tilde{x}}^\alpha \eta_{\alpha\beta} \dot{\tilde{x}}^\beta=0
\ee
and the velocity of propagation in the generalized space-time  is energy independent. 

For the second choice of space-time  coordinates, one has
\be
\varphi^\alpha_\mu(p) = \lim_{l\to 0} \frac{\partial(p\oplus l)_\mu}{\partial l_\alpha}, \ee
and using again the composition law in (\ref{DRK1}), one has in this case
\be
\varphi^0_0=1-\frac{p_0}{\Lambda},\quad \varphi_0^i=0,\quad \varphi_i^0=0,\quad \varphi^j_i=\delta_i^j \left(1-\frac{p_0}{\Lambda}\right).
\ee

The velocity of propagation in this second choice of space-time  coordinates turns out to be also energy independent. 

We see then that this characteristic fact of SR is also a property of the two simple choices of generalized space-time  obtained from the locality of interactions.

\section{Complementarity of the Algebraic, Geometric, and Locality Perspectives}
\label{sec:complementarity}

In this section, we explore further the connection between the geometric approach and the algebraic and locality perspectives. 

In the locality perspective, we showed that, with the canonical phase-space coordinates of a particle ($x$, $k$), one can identify generalized space-time coordinates $\tilde{x}^\alpha = x^\mu \varphi^\alpha_\mu(k)$, for which interactions defined by an associative DCL ($\oplus$) of momenta become local. There are two alternative choices (\eqref{eq:phi-magic-L} and \eqref{eq:phi-magic-R}) for the functions $\varphi$ defining the generalized space-time  coordinates:
\be
\varphi^\alpha_\mu(k) \,=\, \lim_{\ell\to 0} \frac{\partial(\ell\oplus k)_\mu}{\partial\ell_\alpha}, \quad\quad\quad
\varphi^\alpha_\mu(k) \,=\, \lim_{\ell\to 0} \frac{\partial(k\oplus \ell)_\mu}{\partial\ell_\alpha}.
\label{alternatives-phi}
\ee

From a geometric perspective, the DCL ($\oplus$) can be used to define two alternatives for translations with parameters $a$ in a curved momentum space with coordinates $k$:
\be
T_a(k) = (a\oplus k),  \quad\quad\quad T_a(k) = (k\oplus a).
\label{alternatives}
\ee

The isometries (leaving one point in momentum space invariant) of the metric constructed from a tetrad invariant under the translation defined by the DCL can be identified with the (nonlinear) Lorentz transformations of the momentum $k$. These transformations together with the energy--momentum relation defined from the squared distance between the origin and a point with coordinates $k$ in momentum space and the nonlinear energy--momentum conservation law associated with the DCL define an RDK. 

The translations defined as the first and the second cases in Equation~\eqref{alternatives} have, respectively, the infinitesimal generators
\be
T^\alpha=x^\mu\lim_{l\to 0}\frac{\partial(l\oplus k)_\mu}{\partial l_\alpha} \,,\quad\quad\quad
\bar{T}^\alpha=x^\mu\lim_{l\to 0}\frac{\partial(k\oplus l)_\mu}{\partial l_\alpha}.
\label{two-generators}
\ee

Compared with Equation~\eqref{alternatives-phi}, we see that the two alternatives to introducing generalized space-time coordinates correspond, from the geometric perspective,
to taking them as the generators of the two alternatives for translations that can be associated with a given composition law.

One can alternatively start from a tetrad in a maximally symmetric curved momentum space and identify the translation $T_a(k)$ that leaves it invariant. With this translation, one has two alternatives to define an associative DCL:
\be
(a\oplus k) = T_a(k), \quad\quad\quad  (k\oplus a) = T_a(k),
\ee
which together with the isometries (leaving the origin invariant) identified as nonlinear Lorentz transformations and the energy--momentum relation defined from the squared distance between the origin and a point with coordinates $k$ in momentum space define an RDK. 
Taking the first option by convention, as we explained after Equation~\eqref{T-CL}, the $T^\alpha$ and $\bar{T}^\alpha$ in Equation~\eqref{two-generators} are then seen, respectively, as the generators of translations and of another set of transformations obtained from the (inverse of the) tetrad in momentum~space:
\be
T^\alpha=x^\mu \mathcal{T}^\alpha_\mu (k) \,,\quad\quad\quad
\bar{T}^\alpha=x^\mu e^\alpha_\mu (k).
\ee

From this point of view, the two alternatives for the generalized space-time coordinates correspond either to the 
generators of translations in momentum space or to linear combinations of the canonical space-time  coordinates with the (inverse of the) tetrad in momentum space as coefficients.

The two perspectives, the one based on the locality of interactions and the one based on the geometry of a curved momentum space, provide then two complementary interpretations of the possibility to go beyond special relativity, introducing a new energy scale while maintaining the relativity principle.

We worked out explicitly the relationship between an RDK, and the geometric and the locality approaches in the case of an associative composition law, since, as we have seen, for any associative composition law, it is possible to define space-time  coordinates in which the interaction between two particles is local. The relativistic deformed kinematics that is derived in this case is $\kappa$-Poincaré kinematics, which stands then as the only kinematics derived from the geometric approach that is compatible with locality.

There is a still a third perspective, based on the implementation of symmetries in a noncommutative space-time  (Hopf algebras). In the case of $\kappa$-Poincaré, the space-time coordinates together with the Lorentz generators close the Lie algebra~\cite{KowalskiGlikman:2002jr}:
\be
\lbrace x^0, x^i \rbrace \,=\, \frac{1}{\Lambda} x^i, \qquad \lbrace x^0, J^{0i} \rbrace \,=\, x^i + \frac{1}{\Lambda} J^{0i} ,\qquad \lbrace x^j, J^{0i}\rbrace \,=\, \delta^i_j x^0 + \frac{1}{\Lambda} J^{ji}.
\label{eq:boost_pb}
\ee

In fact, this is the algebra of the generators of isometries in the geometric approach if one identifies the generators of translations with the space-time  coordinates in the algebraic approach. The Lorentz transformations of the one- and two-particle systems in the algebraic approach are defined by the ($\kappa$-)deformation of the Poincaré algebra and the coproduct of the Lorentz generators. It has been shown~\cite{Carmona:2019fwf,Carmona:2019vsh} that they coincide with the Lorentz transformations derived in the geometric approach. We see in this way that this third algebraic perspective is compatible with perspectives based on the geometry of a curved moment space or on the locality of interactions. 

In summary, the main result of this work is the identification of a relation between the geometry of a curved momentum space, the loss of absolute locality, and a relativistic deformed kinematics, through the identification of a common ingredient: a deformed composition law of momenta. While the physical consequences of this result are an open question, the relationship between elements that are usually considered in different approaches to quantum gravity is intriguing and this work may be seen as a first step for future work in this direction.

\authorcontributions{All authors contributed equally to the present work.}

\acknowledgments{This work is supported by Spanish grants PGC2018-095328-B-I00 (FEDER/Agencia estatal de investigación), and DGIID-DGA No. 2015-E24/2. The authors would like to acknowledge the contribution of the COST Action CA18108 ``Quantum gravity phenomenology in the multi-messenger approach''.}

\conflictsofinterest{The authors declare no conflict of interest.} 







\externalbibliography{no}



\begin{thebibliography}{999}

\bibitem[Born(1938)]{Born:1938}
Born, M.
\newblock A Suggestion for Unifying Quantum Theory and Relativity.
\newblock {\em Proc. R. Soc. Lond. Ser. A
  Math. Phys. Sci.} {\bf 1938}, {\em 165},~291--303.
\newblock
  doi:{\changeurlcolor{black}\href{https://doi.org/10.1098/rspa.1938.0060}{\detokenize{10.1098/rspa.1938.0060}}}.

\bibitem[Snyder(1947)]{Snyder:1946qz}
Snyder, H.S.
\newblock {Quantized space-time}.
\newblock {\em Phys. Rev.} {\bf 1947}, {\em 71},~38--41,
\newblock
  doi:{\changeurlcolor{black}\href{https://doi.org/10.1103/PhysRev.71.38}{\detokenize{10.1103/PhysRev.71.38}}}.

\bibitem[Mukhi(2011)]{Mukhi:2011zz}
Mukhi, S.
\newblock {String theory: A perspective over the last 25 years}.
\newblock {\em Class. Quantum Gravity} {\bf 2011}, {\em 28},~153001,
  \href{http://arxiv.org/abs/1110.2569}{{\normalfont
  [arXiv:physics.pop-ph/1110.2569]}}.
\newblock
  doi:{\changeurlcolor{black}\href{https://doi.org/10.1088/0264-9381/28/15/153001}{\detokenize{10.1088/0264-9381/28/15/153001}}}.

\bibitem[Aharony(2000)]{Aharony:1999ks}
Aharony, O.
\newblock {A Brief review of 'little string theories'}.
\newblock {\em Class. Quantum Gravity} {\bf 2000}, {\em 17},~929--938,
  \href{http://arxiv.org/abs/hep-th/9911147}{{\normalfont
  [arXiv:hep-th/hep-th/9911147]}}.
\newblock
  doi:{\changeurlcolor{black}\href{https://doi.org/10.1088/0264-9381/17/5/302}{\detokenize{10.1088/0264-9381/17/5/302}}}.

\bibitem[Dienes(1997)]{Dienes:1996du}
Dienes, K.R.
\newblock {String theory and the path to unification: A Review of recent
  developments}.
\newblock {\em Phys. Rept.} {\bf 1997}, {\em 287},~447--525,
  \href{http://arxiv.org/abs/hep-th/9602045}{{\normalfont
  [arXiv:hep-th/hep-th/9602045]}}.
\newblock
  doi:{\changeurlcolor{black}\href{https://doi.org/10.1016/S0370-1573(97)00009-4}{\detokenize{10.1016/S0370-1573(97)00009-4}}}.

\bibitem[Sahlmann(2010)]{Sahlmann:2010zf}
Sahlmann, H.
\newblock {Loop Quantum Gravity---A Short Review}.
\newblock  {In Proceedings of the Foundations of Space and Time: Reflections on Quantum
  Gravity, Cape Town, South Africa}, 10-14 August 2009, 
  \href{http://arxiv.org/abs/1001.4188}{{\normalfont
  [arXiv:gr-qc/1001.4188]}}.

\bibitem[Dupuis \em{et~al.}(2012)Dupuis, Ryan, and Speziale]{Dupuis:2012yw}
Dupuis, M.; Ryan, J.P.; Speziale, S.
\newblock {Discrete gravity models and Loop Quantum Gravity: A short review}.
\newblock {\em SIGMA} {\bf 2012}, {\em 8},~052,
  \href{http://arxiv.org/abs/1204.5394}{{\normalfont
  [arXiv:gr-qc/1204.5394]}}.
\newblock
  doi:{\changeurlcolor{black}\href{https://doi.org/10.3842/SIGMA.2012.052}{\detokenize{10.3842/SIGMA.2012.052}}}.

\bibitem[Van~Nieuwenhuizen(1981)]{VanNieuwenhuizen:1981ae}
Van~Nieuwenhuizen, P.
\newblock {Supergravity}.
\newblock {\em Phys. Rept.} {\bf 1981}, {\em 68},~189--398,
\newblock
  doi:{\changeurlcolor{black}\href{https://doi.org/10.1016/0370-1573(81)90157-5}{\detokenize{10.1016/0370-1573(81)90157-5}}}.

\bibitem[Taylor(1984)]{Taylor:1983su}
Taylor, J.G.
\newblock {A Review of Supersymmetry and Supergravity}.
\newblock {\em Prog. Part. Nucl. Phys.} {\bf 1984}, {\em 12},~1--101,
\newblock
  doi:{\changeurlcolor{black}\href{https://doi.org/10.1016/0146-6410(84)90002-4}{\detokenize{10.1016/0146-6410(84)90002-4}}}.

\bibitem[Wallden(2010)]{Wallden:2010sh}
Wallden, P.
\newblock {Causal Sets: Quantum Gravity from a Fundamentally Discrete
  Spacetime}.
\newblock {\em J. Phys. Conf. Ser.} {\bf 2010}, {\em 222},~012053,
  \href{http://arxiv.org/abs/1001.4041}{{\normalfont
  [arXiv:gr-qc/1001.4041]}}.
\newblock
  doi:{\changeurlcolor{black}\href{https://doi.org/10.1088/1742-6596/222/1/012053}{\detokenize{10.1088/1742-6596/222/1/012053}}}.

\bibitem[Wallden(2013)]{Wallden:2013kka}
Wallden, P.
\newblock {Causal Sets Dynamics: Review \& Outlook}.
\newblock {\em J. Phys. Conf. Ser.} {\bf 2013}, {\em 453},~012023,
\newblock
  doi:{\changeurlcolor{black}\href{https://doi.org/10.1088/1742-6596/453/1/012023}{\detokenize{10.1088/1742-6596/453/1/012023}}}.

\bibitem[Henson(2009)]{Henson:2006kf}
Henson, J.
\newblock {The Causal set approach to quantum gravity}. In {\em Approaches to
  Quantum Gravity: Toward a New Understanding of Space, Time and Matter};
  Oriti, D., Ed.; Cambridge University Press:  Cambridge, UK,  2009; pp. 393--413. 
  \href{http://arxiv.org/abs/gr-qc/0601121}{{\normalfont
  [arXiv:gr-qc/gr-qc/0601121]}}.

\bibitem[Amelino-Camelia(2013)]{AmelinoCamelia:2008qg}
Amelino-Camelia, G.
\newblock {Quantum-Spacetime Phenomenology}.
\newblock {\em Living Rev. Relativ.} {\bf 2013}, {\em 16},~5,
  \href{http://arxiv.org/abs/0806.0339}{{\normalfont
  [arXiv:gr-qc/0806.0339]}}.
\newblock
  doi:{\changeurlcolor{black}\href{https://doi.org/10.12942/lrr-2013-5}{\detokenize{10.12942/lrr-2013-5}}}.

\bibitem[Majid(1995)]{Majid:1995qg}
Majid, S.
\newblock {\em Foundations of Quantum Group Theory}; Cambridge University
  Press:  Cambridge, UK, 1995.

\bibitem[Majid and Ruegg(1994)]{Majid1994}
Majid, S.; Ruegg, H.
\newblock {Bicrossproduct structure of kappa Poincare group and noncommutative
  geometry}.
\newblock {\em Phys. Lett.} {\bf 1994}, {\em B334},~348--354,
  \href{http://arxiv.org/abs/hep-th/9405107}{{\normalfont
  [arXiv:hep-th/hep-th/9405107]}}.
\newblock
  doi:{\changeurlcolor{black}\href{https://doi.org/10.1016/0370-2693(94)90699-8}{\detokenize{10.1016/0370-2693(94)90699-8}}}.

\bibitem[Lukierski \em{et~al.}(1995)Lukierski, Ruegg, and
  Zakrzewski]{Lukierski1995}
Lukierski, J.; Ruegg, H.; Zakrzewski, W.J.
\newblock {Classical quantum mechanics of free kappa relativistic systems}.
\newblock {\em Ann. Phys.} {\bf 1995}, {\em 243},~90--116,
  \href{http://arxiv.org/abs/hep-th/9312153}{{\normalfont
  [arXiv:hep-th/hep-th/9312153]}}.
\newblock
  doi:{\changeurlcolor{black}\href{https://doi.org/10.1006/aphy.1995.1092}{\detokenize{10.1006/aphy.1995.1092}}}.

\bibitem[Kowalski-Glikman(2002)]{KowalskiGlikman:2002ft}
Kowalski-Glikman, J.
\newblock {De sitter space as an arena for doubly special relativity}.
\newblock {\em Phys. Lett.} {\bf 2002}, {\em B547},~291--296,
  \href{http://arxiv.org/abs/hep-th/0207279}{{\normalfont
  [arXiv:hep-th/hep-th/0207279]}}.
\newblock
  doi:{\changeurlcolor{black}\href{https://doi.org/10.1016/S0370-2693(02)02762-4}{\detokenize{10.1016/S0370-2693(02)02762-4}}}.

\bibitem[Amelino-Camelia \em{et~al.}(2011)Amelino-Camelia, Freidel,
  Kowalski-Glikman, and Smolin]{AmelinoCamelia:2011bm}
Amelino-Camelia, G.; Freidel, L.; Kowalski-Glikman, J.; Smolin, L.
\newblock {The principle of relative locality}.
\newblock {\em Phys. Rev.} {\bf 2011}, {\em D84},~084010,
  \href{http://arxiv.org/abs/1101.0931}{{\normalfont
  [arXiv:hep-th/1101.0931]}}.
\newblock
  doi:{\changeurlcolor{black}\href{https://doi.org/10.1103/PhysRevD.84.084010}{\detokenize{10.1103/PhysRevD.84.084010}}}.

\bibitem[Lobo and Palmisano(2016)]{Lobo:2016blj}
Lobo, I.P.; Palmisano, G.
\newblock {Geometric interpretation of Planck-scale-deformed co-products}.
\newblock {\em Int. J. Mod. Phys. Conf. Ser.} {\bf 2016}, {\em 41},~1660126,
  \href{http://arxiv.org/abs/1612.00326}{{\normalfont
  [arXiv:hep-th/1612.00326]}}.
\newblock
  doi:{\changeurlcolor{black}\href{https://doi.org/10.1142/S2010194516601265}{\detokenize{10.1142/S2010194516601265}}}.

\bibitem[Kowalski-Glikman and Nowak(2003)]{KowalskiGlikman:2002jr}
Kowalski-Glikman, J.; Nowak, S.
\newblock {Noncommutative space-time of doubly special relativity theories}.
\newblock {\em Int. J. Mod. Phys.} {\bf 2003}, {\em D12},~299--316,
  \href{http://arxiv.org/abs/hep-th/0204245}{{\normalfont
  [arXiv:hep-th/hep-th/0204245]}}.
\newblock
  doi:{\changeurlcolor{black}\href{https://doi.org/10.1142/S0218271803003050}{\detokenize{10.1142/S0218271803003050}}}.

\bibitem[Carmona \em{et~al.}(2019)Carmona, Cortés, and
  Relancio]{Carmona:2019fwf}
Carmona, J.M.; Cortés, J.L.; Relancio, J.J.
\newblock {Relativistic deformed kinematics from momentum space geometry}.
\newblock {\em Phys. Rev.} {\bf 2019}, {\em D100},~104031,
  \href{http://arxiv.org/abs/1907.12298}{{\normalfont
  [arXiv:hep-th/1907.12298]}}.
\newblock
  doi:{\changeurlcolor{black}\href{https://doi.org/10.1103/PhysRevD.100.104031}{\detokenize{10.1103/PhysRevD.100.104031}}}.

\bibitem[Carmona \em{et~al.}(2018)Carmona, Cortes, and
  Relancio]{Carmona:2017cry}
Carmona, J.M.; Cortes, J.L.; Relancio, J.J.
\newblock {Spacetime from locality of interactions in deformations of special
  relativity: The example of $\kappa$-Poincar\'e Hopf algebra}.
\newblock {\em Phys. Rev.} {\bf 2018}, {\em D97},~064025,
  \href{http://arxiv.org/abs/1711.08403}{{\normalfont
  [arXiv:hep-th/1711.08403]}}.
\newblock
  doi:{\changeurlcolor{black}\href{https://doi.org/10.1103/PhysRevD.97.064025}{\detokenize{10.1103/PhysRevD.97.064025}}}.

\bibitem[Carmona \em{et~al.}(2020)Carmona, Cortés, and
  Relancio]{Carmona:2019vsh}
Carmona, J.M.; Cortés, J.L.; Relancio, J.J.
\newblock {Relativistic deformed kinematics from locality conditions in a
  generalized spacetime}.
\newblock {\em Phys. Rev.} {\bf 2020}, {\em D101},~044057,
  \href{http://arxiv.org/abs/1912.12885}{{\normalfont
  [arXiv:hep-th/1912.12885]}}.
\newblock
  doi:{\changeurlcolor{black}\href{https://doi.org/10.1103/PhysRevD.101.044057}{\detokenize{10.1103/PhysRevD.101.044057}}}.

\bibitem[Carmona \em{et~al.}(2012)Carmona, Cortes, and Mercati]{Carmona:2012un}
Carmona, J.; Cortes, J.; Mercati, F.
\newblock {Relativistic kinematics beyond Special Relativity}.
\newblock {\em Phys.\ Rev.\ D} {\bf 2012}, {\em 86},~084032,
  \href{http://arxiv.org/abs/1206.5961}{{\normalfont
  [arXiv:hep-th/1206.5961]}}.
doi:{\changeurlcolor{black}\href{https://doi.org/10.1103/PhysRevD.86.084032}{\detokenize{10.1103/PhysRevD.86.084032}}}.

\bibitem[Carmona \em{et~al.}(2016)Carmona, Cortes, and
  Relancio]{Carmona:2016obd}
Carmona, J.; Cortes, J.; Relancio, J.
\newblock {Beyond Special Relativity at second order}.
\newblock {\em Phys.\ Rev.\ D} {\bf 2016}, {\em 94},~084008,\linebreak 
  \href{http://arxiv.org/abs/1609.01347}{{\normalfont
  [arXiv:hep-th/1609.01347]}}.
\newblock
  doi:{\changeurlcolor{black}\href{https://doi.org/10.1103/PhysRevD.94.084008}{\detokenize{10.1103/PhysRevD.94.084008}}}.

\bibitem[Kowalski-Glikman and Nowak(2002)]{KowalskiGlikman:2002we}
Kowalski-Glikman, J.; Nowak, S.
\newblock {Doubly special relativity theories as different bases of kappa
  Poincare algebra}.
\newblock {\em Phys. Lett.} {\bf 2002}, {\em B539},~126--132,
  \href{http://arxiv.org/abs/hep-th/0203040}{{\normalfont
  [arXiv:hep-th/hep-th/0203040]}}.
\newblock
  doi:{\changeurlcolor{black}\href{https://doi.org/10.1016/S0370-2693(02)02063-4}{\detokenize{10.1016/S0370-2693(02)02063-4}}}.

\bibitem[Battisti and Meljanac(2010)]{Battisti:2010sr}
Battisti, M.V.; Meljanac, S.
\newblock {Scalar Field Theory on Non-commutative Snyder Space-Time}.
\newblock {\em Phys. Rev.} {\bf 2010}, {\em D82},~024028,
  \href{http://arxiv.org/abs/1003.2108}{{\normalfont
  [arXiv:hep-th/1003.2108]}}.
\newblock
  doi:{\changeurlcolor{black}\href{https://doi.org/10.1103/PhysRevD.82.024028}{\detokenize{10.1103/PhysRevD.82.024028}}}.

\bibitem[Meljanac \em{et~al.}(2009)Meljanac, Meljanac, Samsarov, and
  Stojic]{Meljanac:2009ej}
Meljanac, S.; Meljanac, D.; Samsarov, A.; Stojic, M.
\newblock {Lie algebraic deformations of Minkowski space with Poincare algebra}. \emph{arXiv} {\bf 2009}, arXiv:0909.1706.
 \href{http://arxiv.org/abs/0909.1706}{{\normalfont
  [arXiv:math-ph/0909.1706]}}.

\bibitem[Chern \em{et~al.}(1999)Chern, Chen, and Lam]{Chern:1999jn}
Chern, S.S.; Chen, W.H.; Lam, K.S.
\newblock {\em {Lectures on Differential Geometry}}; World Scientific: Singapore,
  1999; see Eqs. (1.30) and (1.31) of Chapter 6. 

\bibitem[Gubitosi and Mercati(2013)]{Gubitosi:2013rna}
Gubitosi, G.; Mercati, F.
\newblock {Relative Locality in $\kappa$-Poincar\'e}.
\newblock {\em Class. Quantum Gravity} {\bf 2013}, {\em 30},~145002,
 \href{http://arxiv.org/abs/1106.5710}{{\normalfont
  [arXiv:gr-qc/1106.5710]}}.
\newblock
  doi:{\changeurlcolor{black}\href{https://doi.org/10.1088/0264-9381/30/14/145002}{\detokenize{10.1088/0264-9381/30/14/145002}}}.

\bibitem[Relancio and Liberati(2020)]{Relancio:2020zok}
Relancio, J.J.; Liberati, S.
\newblock {Phenomenological consequences of a geometry in the cotangent
  bundle}.
\newblock {\em Phys. Rev.} {\bf 2020}, {\em D101},~064062,
  \href{http://arxiv.org/abs/2002.10833}{{\normalfont
  [arXiv:gr-qc/2002.10833]}}.
\newblock
  doi:{\changeurlcolor{black}\href{https://doi.org/10.1103/PhysRevD.101.064062}{\detokenize{10.1103/PhysRevD.101.064062}}}.

\end{thebibliography}
\end{document}